\documentclass[aps,prb,twocolumn,nobibnotes]{revtex4-2}  
\usepackage{dcolumn}
\usepackage{graphicx}
\usepackage{color}
\usepackage{amssymb}
\usepackage{amsmath}
\usepackage{bm}
\usepackage[colorlinks=true,linktocpage=true,breaklinks=true]{hyperref}
\usepackage{multirow}
\usepackage{braket}
\usepackage{appendix}

\begin{document}
\title{Efficient Generation of Spin Currents in Altermagnets via Magnon Drag}
\author{Konstantinos Sourounis}
\email{konstantinos.sourounis@univ-amu.fr}
\author{Aurélien Manchon}
\email{aurelien.manchon@univ-amu.fr}
\affiliation{Aix-Marseille Université, CNRS, CINaM, Marseille, France}
\begin{abstract}
Altermagnets, a recently identified class of magnetic materials, possess a spin-split Fermi surface that results in the so-called spin splitter effect, enabling the generation of a spin current transverse to the injection direction and whose polarization lies along the Néel vector. In this study, we investigate how magnons interact with electrons in an altermagnetic metal. We find that while the electron-magnon interaction does not perturb the magnon dispersion, a charge current flowing in the material can induce a transverse magnon spin current, analogous to the electronic spin splitter effect. This spin current possesses both electronic and magnonic characteristics, i.e., a chemical potential dependence and a strong temperature dependence. This effect realizes the efficient generation of spin currents via magnons without depending on the material's spin-orbit coupling.
\end{abstract}

\maketitle

\section{INTRODUCTION} \label{sec:i}

Spin currents are the workhorse of spintronics, enabling the transport of angular momentum over long distances \cite{Tombros2007}, the current-driven excitation and switching of magnetic materials \cite{Ralph2008,Manchon2019}, the interconversion between charge and spin mediated by spin-orbit coupling via the spin Hall effect (SHE) \cite{sinova2015spin} and the spin Rashba-Edelstein effect \cite{Bihlmayer2022}, and the introduction of new device concepts \cite{Choi2018,Manipatruni2019,Noel2020}. Whereas most of these phenomena harvest electronic spin currents, magnon propagation in insulating ferro- and ferrimagnets are attracting increasing attention due to their ability for high-frequency information processing and low-power dissipation \cite{Serga2010yig,Chumak2015magnon}. Remarkably, magnonic spin currents possess numerous features akin to electronic spin currents such as long-range propagation \cite{cornelissen2015long}, spin-valve behavior \cite{Wu2018,Chen2021,Li2024} but also magnonic spin transfer \cite{Wang2019, Zheng2022} and spin-orbit torques \cite{Manchon2014a,Kovalev2016,Kim2019f}. In magnetic metals, electrons and magnons interact strongly, unlocking the magnon-drag effect \cite{Blatt1967,Costache2012,Watzman2016}, i.e., the dragging of electronic current by a magnonic flux (and vice-versa), recently extended to heterostructures involving magnetic insulators \cite{Zhang2012d,li2016observation,wu2016observation,cheng2017interplay}.

Whereas most progress in electronic and magnonic spintronics has been achieved in materials and heterostructures involving ferro- and ferrimagnets, the realization that antiferromagnets can play an active role in spin transport setups \cite{jungwirth2016antiferromagnetic,baltz2018antiferromagnetic}, experiencing spin torque \cite{Zelezny2014,Wadley2015b,Tsai2020} and enabling spin pumping \cite{Vaidya2020,Li2020c}, has severely expanded the horizon of spintronics. Remarkably, since antiferromagnets display a vast range of magnetic orders \cite{Bonbien2022} (collinear, coplanar, non-collinear, with inversion, mirror or rotation symmetry breaking, etc.), the spin-charge interconversion processes can adopt unconventional symmetries depending on the magnetic space group of the material. In conventional collinear antiferromagnets, the interconversion between charge and spin currents driven by SHE adopts the same symmetry as in nonmagnetic metals \cite{Zhang2014e}, i.e., the spin polarization is locked perpendicular to the plane defined by the currents propagation directions. However, antiferromagnets displaying a noncollinear magnetic order such as Mn$_3$X (X=Ga, Ge, Sn) display non-relativistic spin Hall \cite{Zhang2018d} and Rashba-Edelstein effects \cite{Gonzalez-Hernandez2024} (i.e., even in the absence of spin-orbit coupling), as well as the unusual magnetic spin Hall effect (MSHE) \cite{Zelezny2017b,Kimata2019}. This phenomenon exhibits two distinct features compared to conventional SHE: first, the polarization of the spin Hall current induced by the MSHE is governed by the magnetic configuration and, in Mn$_3$X, lies in the plane of the magnetic moments; second, this effect is odd under time-reversal symmetry and therefore changes sign upon reversing the magnetic moments. The MSHE also exists in collinear antiferromagnets as long as the parity-time symmetry is broken \cite{Chen2020c} (e.g., Mn$_2$Au) or when the two sublattices are related by a rotation \cite{gonzalez2021efficient}. In these two cases, the polarization of the spin Hall current lies along the Néel vector. It is important to emphasize that although this mechanism has been called "antiferromagnetic spin Hall effect" \cite{Chen2020c} and "spin-splitter effect" \cite{gonzalez2021efficient}, respectively, they are fundamentally the same effect as the MSHE, although in a different magnetic landscape. Recent reports have shown that the MSHE-driven spin currents can be used to generate spin torque in neighboring magnets \cite{Nan2020,bose2022tilted,bai2022observation,karube2022observation}. The transport properties of magnons in antiferromagnets are currently actively studied, with the demonstration of long-range propagation \cite{lebrun2018tunable,xing2019magnon}, the prediction of the analog of SHE \cite{Zyuzin2016b,Cheng2016b} and the analog of the Rashba-Edelstein effect \cite{Li2020d}, and the investigation of the interplay between magnonic and electronic spin currents \cite{Wen2019,cheng2020spin,erlandsen2022magnon,barbeau2023nonequilibrium}. 

\begin{figure}[b]
\includegraphics[width=8.6cm,height=8.6cm,keepaspectratio]{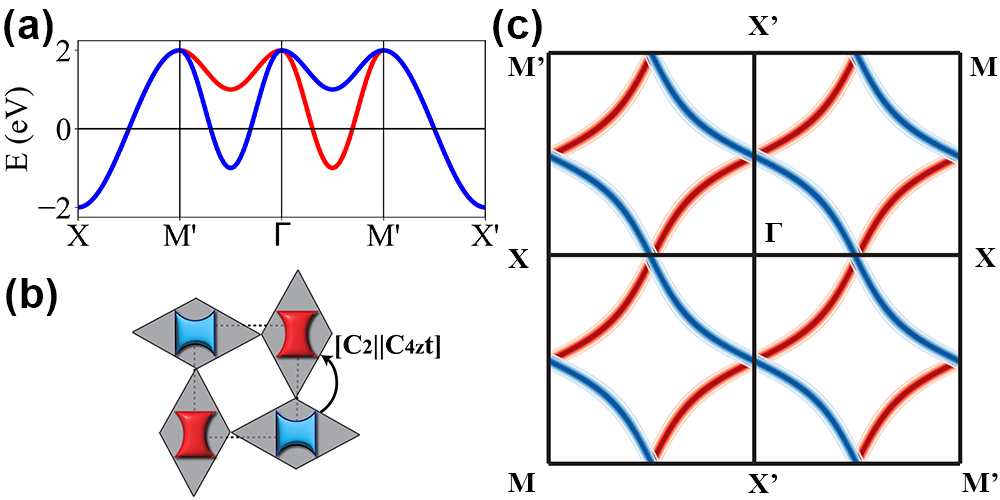}
\caption{(a) The energy spectrum of the altermagnetic electrons along the high-symmetry path. (b) In the $d$-wave altermagnet considered, the electron orbitals are related to each other by a crystal and spin rotation. (c) The Fermi surface of the electronic model at $E=0$.}
\label{fig:fig1}
\end{figure}

In the present work, we focus on the electron-magnon interaction in collinear antiferromagnets whose sublattices are connected by a rotation operation, see Fig. \ref{fig:fig1}, featuring a distinct Type-III spin symmetry class of magnetism with compensated even-partial wave spin order. This class of magnetic materials, named altermagnets \cite{smejkal2022prxb}, display momentum-symmetric spin-splitting \cite{SLM12,Noda16,Okugawa18,Smejkal2020,Naka2019, Hayami2019, Ahn2019,Yuan2020, Yuan2021,vsmejkal2022emerging,Egorov2021a,Guo2023,mazin2021prediction,smejkal2022prxb}, anomalous Hall effect \cite{Smejkal2020,mazin2021prediction,feng2022anomalous,gonzalez2023spontaneous,reichlova2024observation} and electronic MSHE \cite{gonzalez2021efficient,bose2022tilted,bai2022observation,karube2022observation}. Experiments based on the MSHE have been realized in $\rm{RuO_2}$  with the generation of spin currents and torques \cite{bose2022tilted,bai2022observation,karube2022observation}. Debates on the exact nature of the magnetism in these materials exist \cite{smolyanyuk2024fragility}; however, it is expected that altermagnetism is sourced in the itinerant electrons of the metal \cite{berlijn2017itinerant,zhu2019anomalous,lovesey2023magnetic,smolyanyuk2024fragility}. Interestingly, peculiar symmetries of altermagnetic crystals also impact the magnon dispersion. In conventional bipartite collinear antiferromagnets, the two magnon modes are degenerate. In altermagnets, this degeneracy is lifted along the low symmetry directions of the crystals \cite{vsmejkal2022chiral,cui2023efficient,brekke2023two}, favoring the propagation of one given chirality along specific crystallographic directions.

In this work, we show how the itinerant altermagnetic electrons can imprint their "altermagnetic" properties on the magnons via the electron-magnon interaction. At equilibrium, the interaction averages to zero. However, out of equilibrium, the electron-magnon scattering locks the magnon spin current on the electronic spin current, realizing the magnonic MSHE or spin-splitter effect. Compared with the electronic MSHE, the magnonic MSHE exhibits a sizable temperature dependence. In addition, it possesses both electronic and magnonic characteristics, offering knobs to tune the magnonic current in such systems. 

\section{ELECTRON-MAGNON COUPLING} \label{sec:ii}
\subsection{Altermagnetic electrons and antiferromagnetic magnons}
We consider a simple electron model with symmetric momentum-dependent spin-splitting \cite{vsmejkal2022emerging}
\begin{equation}
    \hat{H}^e_{\bm k} = 2t\cos{k_x}\cos{k_y}+2IS\sigma_z\sin{k_x}\sin{k_y}, \label{eq:1}
\end{equation}
where $t$ is the electron hopping and $I$ is the exchange coupling strength between the itinerant electrons and the local magnetic moment. This minimal Hamiltonian model a typically $d$-wave altermagnet, as depicted in Fig. \ref{fig:fig1}(b). The electronic band structure consists in two energy bands $\epsilon_{\boldsymbol{k},\downarrow/\uparrow}=2t\cos{k_x}\cos{k_y}\pm 2IS\sin{k_x}\sin{k_y}$, as displayed in Fig. \ref{fig:fig1}(a). Different bands dominate the transport along different directions of the Brillouin zone, as illustrated by their Fermi surface contour in Fig. \ref{fig:fig1}(c).

The spin Hamiltonian of the square lattice is
\begin{eqnarray}
    \hat{H}^m &=&  J\sum_{\langle ij\rangle} {\bf S}_{A,i}\cdot {\bf S}_{B,j}+K\sum_i(S^z_i)^2, \label{eq:2}
\end{eqnarray}
where $J$ is the Heisenberg exchange, and $K$ is the easy axis anisotropy that determines the gap in the magnon spectrum. Notice that, in principle, in altermagnets, the next-nearest neighbor exchange lifts the degeneracy between the two magnon chiralities, as mentioned above \cite{vsmejkal2022chiral,cui2023efficient,brekke2023two}. Recent reports on MnTe \cite{Liu2024,Jost2025} suggest that this next-nearest neighbor exchange is about 0.06 meV, almost two orders of magnitude smaller than the direct exchange interaction (4 meV in MnTe). Notice that a much larger value, exceeding 1.5 meV, has been computed in the metallic rutile RuO$_2$ \cite{vsmejkal2022chiral}. We nonetheless emphasize that altermagnetism in this compound remains a controversial issue \cite{smolyanyuk2024fragility,Kessler2024}. Hence, we focus our attention on the effect of the electron spin splitting, as accounting for the degeneracy lift does not qualitatively impact our conclusions.

By performing the Holstein-Primakoff transformation to the first order, we get the magnon Hamiltonian $\hat{H}^m_{\boldsymbol{q}}$. To diagonalize this Hamiltonian, we solve \cite{rezende2019introduction} $\hat{T}^{\dagger}_{\boldsymbol{q}}\hat{H}^m_{\boldsymbol{q}}\hat{T}_{\boldsymbol{q}}=\hat{\omega}_{\boldsymbol{q}}$, where the degenerate magnon spectrum is given as $\omega_{\boldsymbol{q},
\pm}=S\sqrt{A_{\boldsymbol{q}}^2-B_{\boldsymbol{q}}^2}$, $A({\bm q})=4J+2K$, $ B({\bm q})= 4J\cos{(q_x/2)}\cos{(q_y/2)}$. Throughout this work, we set $t=1$ (eV), $I=0.5t$, $S=1$, $J=t/100$, and $K=J/10$, unless specified otherwise. 

Finally, we consider the electron-magnon coupling
\begin{eqnarray}
H^{em}_{\bm k}&=&-\sqrt{\frac{I^2S}{2N}}\sum_{\bm q} (a_{\boldsymbol{q}}+b^{\dagger}_{-\boldsymbol{q}})c^{\dagger}_{\boldsymbol{k+q},\downarrow}c_{\boldsymbol{k},\uparrow}\notag \\&+&(a_{-\boldsymbol{q}}^{\dagger}+b_{\boldsymbol{q}})c^{\dagger}_{\boldsymbol{k+q},\uparrow}c_{\boldsymbol{k},\downarrow}+h.c., \label{eq:3}
\end{eqnarray}
where $a_{\bm q}, b_{\bm q}$ are magnon operators relating to the A-(B-) sublattice of the antiferromagnet, which, in the diagonalized basis $\gamma_{{\bm q},\pm}$, is rewritten as
\begin{eqnarray}
H^{em}_{\bm k}&=&-\sqrt{\frac{I^2S}{2N}}\sum_{\bm q} \bigg(W^+_{\bm q}\gamma_{\boldsymbol{q},+}+W^-_{\bm q}\gamma^{\dagger}_{-\boldsymbol{q},-}\bigg)c^{\dagger}_{\boldsymbol{k+q},\downarrow}c_{\boldsymbol{k},\uparrow}\notag\\&+&\bigg(W^+_{\bm q}\gamma^{\dagger}_{\boldsymbol{-q},+}+W^-_{\bm q}\gamma_{\boldsymbol{q},-}\bigg)c^{\dagger}_{\boldsymbol{k+q},\uparrow}c_{\boldsymbol{k},\downarrow}+h.c., \label{eq:4}
\end{eqnarray}
Here, $W^{\nu}_{\bm q}=T_{a,\nu}(\boldsymbol{q})+T_{b,\nu}(\boldsymbol{q})$. $T_{a/b,\nu}(\boldsymbol{q})$ are the elements of the magnon diagonalization matrix $\hat{T}_{\boldsymbol{q}}$ and $\nu=\pm$ corresponds to the magnon chirality.

\subsection{The self-energy of antiferromagnetic magnons}

From the above expression, we calculate the impact of the electron-magnon interaction on the antiferromagnetic magnon spectrum. To do so, we extend the theory of electron-magnon interaction in ferromagnets \cite{woolsey1970electron} to the case of antiferromagnetic magnons. The magnon self-energy becomes
\begin{eqnarray}
    \Pi_{\boldsymbol{q}}^{\nu}(\omega, T)=\frac{(I\sqrt{S}W^{\nu}_{\bm q})^2}{2N}\sum_{\boldsymbol{k}}(\nu)&\times&\notag\\
    \bigg(\frac{f^0_{\boldsymbol{k},\uparrow}-f^0_{\boldsymbol{k+q},\downarrow}}{\omega+i0^++\epsilon_{\boldsymbol{k},\uparrow}-\epsilon_{\boldsymbol{k+q},\downarrow}}
    &+&\frac{f^0_{\boldsymbol{k+q},\uparrow}-f^0_{\boldsymbol{k},\downarrow}}{\omega+i0^++\epsilon_{\boldsymbol{k},\downarrow}-\epsilon_{\boldsymbol{k+q},\uparrow}}\bigg),\notag\\\quad \label{eq:5}
\end{eqnarray}
where $\Pi^+_{\bm q}=-\Pi^-_{\bm q}$ and $f^0_{{\bm k},\updownarrow}$ is the equilibrium distribution of electrons. We emphasize that ${\bm q}$ is the magnon wave vector and ${\bm k}$ is the electron wave vector.

\begin{figure}[t]
\includegraphics[width=8.6cm,height=8.6cm,keepaspectratio]{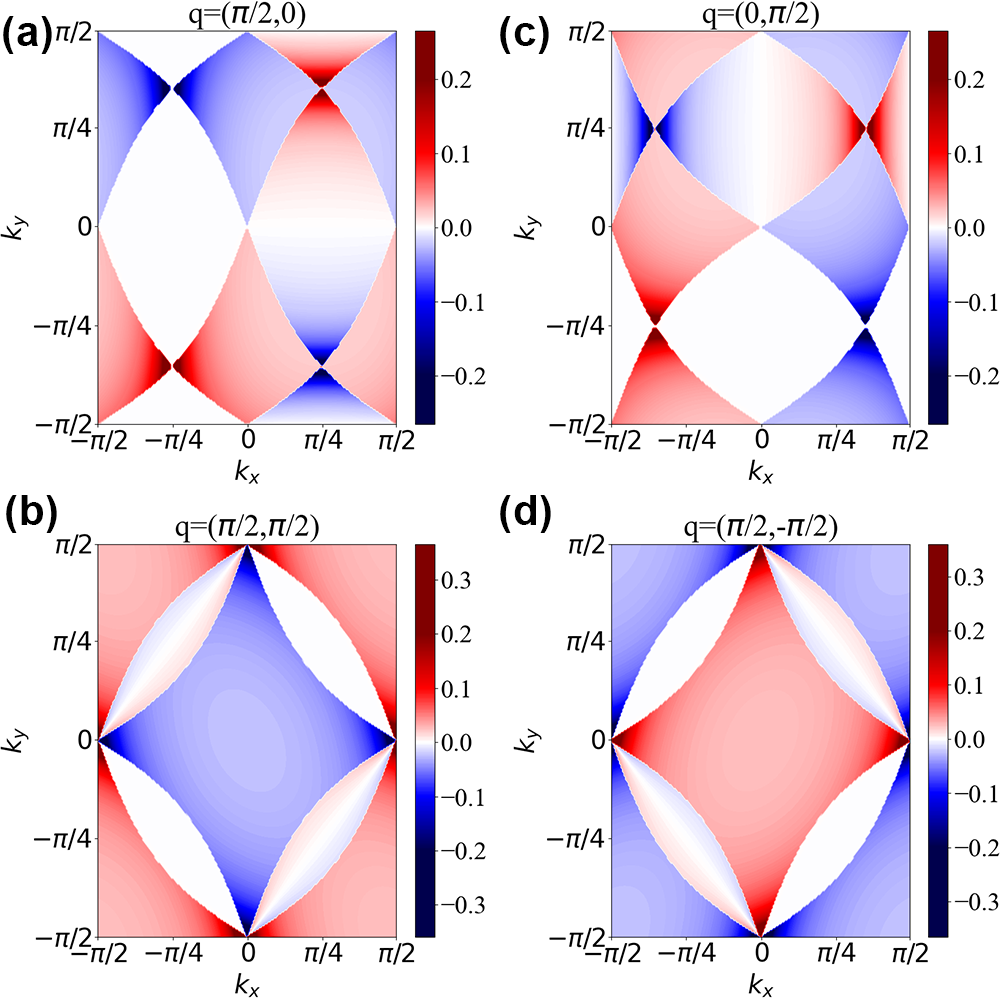}
\caption{The real part of the self-energy of magnons $\Pi_{\boldsymbol{q}}^{+}$ in the k-momentum space of electrons for different momenta of magnons (a) ${\bm q}= (\pi/2,0)$, (b) ${\bm q}= (\pi/2,\pi/2)$, (c) ${\bm q}= (0,\pi/2)$, (d) ${\bm q}= (\pi/2,-\pi/2)$.}
\label{fig:fig2}
\end{figure}

\begin{figure}[t]
\includegraphics[width=8.6cm,height=12.9cm,keepaspectratio]{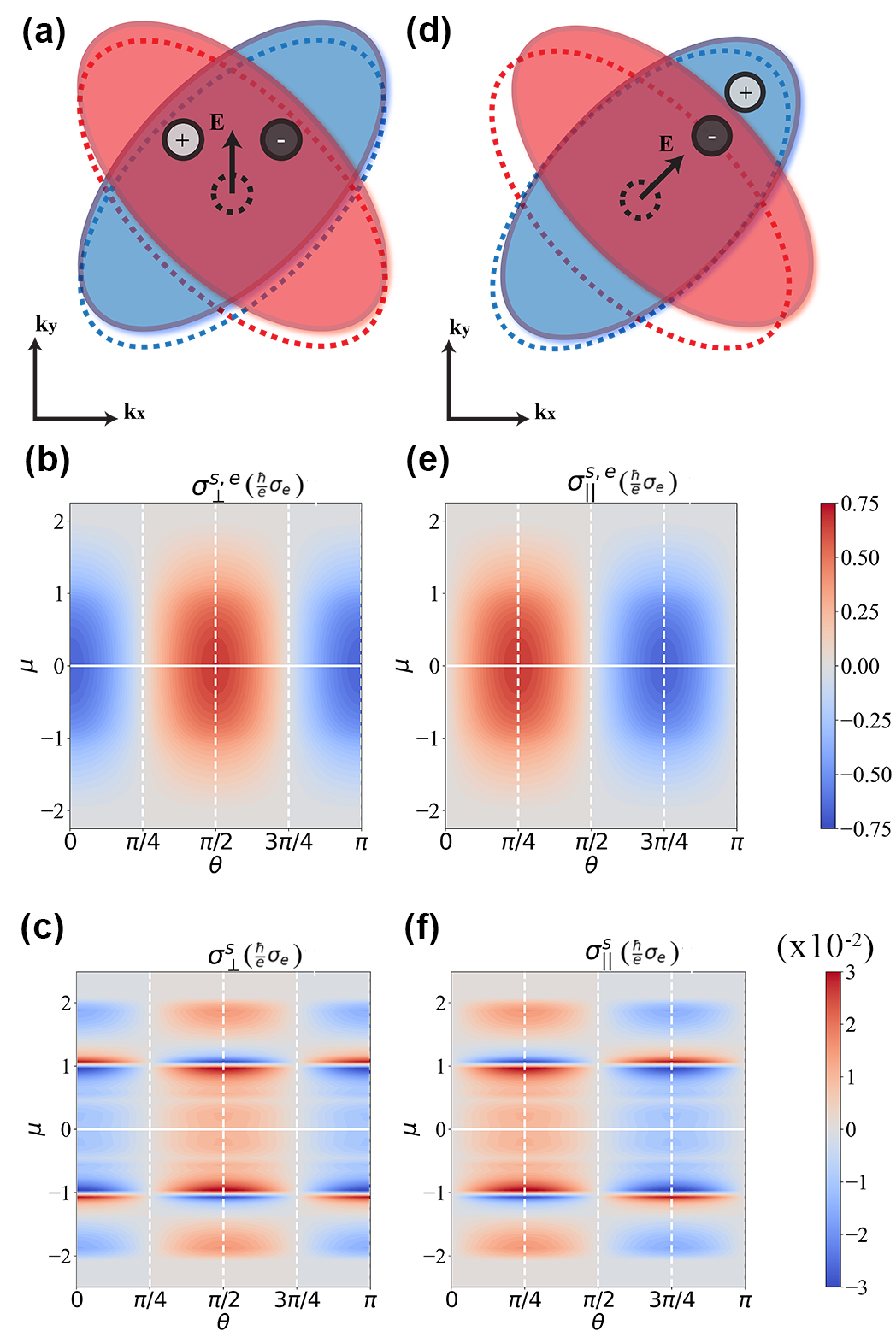}
\caption{(a,d) Schematic representation of the electron-magnon induced spin currents for an electric field at angle (a) $\theta=\pi/2$ (d) $\theta=\pi/4$. The dashed red and blue lines represent the electronic Fermi surface at equilibrium, whereas the magnons are degenerate (dashed black circle). In the presence of an electric field, the Fermi surfaces are shifted, as indicated by solid lines, which drags the two magnon chiralities (solid black circles) either (a) transversally or (d) longitudinally to the electric field. (b,e) The chemical potential and angular dependence of (b) transverse and (e) longitudinal electron spin currents. (c,f) The chemical potential and angular dependence of (c) transverse and (f) longitudinal magnon spin currents. We have set the transport parameters to $T=2J,\tau_m=100\tau_e$.}
\label{fig:fig3}
\end{figure}

In Fig. \ref{fig:fig2}, we calculate the real part of the magnon self-energy in the electron momentum space ${\bm k}$ for four different values of the magnon momentum ${\bm q}$, using the on-shell approximation ($\omega=\omega^\pm_{\bm q}$). We note that when the summation over the Brillouin zone of electrons is performed, then $\Pi_{\boldsymbol{q}}^{\pm}=0$, and the magnon spectrum remains degenerate. In other words, the spin-splitting of the electron band structure is not imprinted on the band structure of magnons. The reason is that the electron-magnon coupling remains zero in equilibrium as the Fermi surfaces of the electrons are balanced. When we calculate the non-equilibrium electron-magnon interaction though, the Fermi surfaces of the two different spin species are imbalanced, depending on the direction of the electric field [see Fig. \ref{fig:fig3} (a) and (d)], and thus the total contribution of the electron-magnon interaction is potentially non-zero.

\subsection{Coupled electron-magnon transport}

The coupled electron-magnon transport is modeled using the Boltzmann transport equation in the relaxation time approximation. The non-equilibrium distribution of electrons with spin $\sigma=\uparrow,\downarrow$ is denoted $f_{\boldsymbol{k},\sigma}$ and obeys the usual Boltzmann equation, with relaxation time $\tau^\sigma$. To the first order in the electric field, 
\begin{eqnarray}
f_{\boldsymbol{k},\sigma}&=&f^0_{\boldsymbol{k},\sigma}-\partial_{\epsilon_{\boldsymbol{k},\sigma}}f^0_{\boldsymbol{k},\sigma }e\hbar\tau^{\sigma}\bigg({\bf v}_k^{\sigma}\cdot \bf{E}\bigg), \label{eq:8}
\end{eqnarray}
where ${\bm\upsilon}_k^{\sigma}$ is the electron velocity and $f_{\boldsymbol{k},\sigma}^0$ is the equilibrium Fermi-Dirac distribution. The Boltzamnn transport equation for magnon in the relaxation time approximation reads \cite{cheng2020spin}
\begin{equation}
    \frac{\partial n_{\boldsymbol{q},\nu}}{\partial t}=\frac{n_{\boldsymbol{q},\nu}-n^{0}_{\boldsymbol{q},\nu}}{\tau_{\nu}}, \label{eq:6}
\end{equation}
where $n_{\boldsymbol{q}}$ is the nonequilibrium distribution of magnons with chirality $\nu$ and $\tau_\nu$ is the associated relaxation time. The above equation is valid as long as the scattering times of the individual magnon chiralities are much longer than the chirality mixing scattering times. Again, the non-equilibrium magnon distribution can be expressed as
\begin{eqnarray}
n_{\boldsymbol{q},\nu}&=&n^{0}_{\boldsymbol{q},\nu}+\partial_{\omega_{\boldsymbol{q},\nu}}n^{0}_{\boldsymbol{q},\nu}g_{\bm q}^{\nu},\label{eq:7}
\end{eqnarray}
where $n^{0}_{\boldsymbol{q},\nu}$ is the equilibrium magnon distribution. The scattering rate of magnons can be derived by applying Fermi's golden rule on the electron-magnon interaction, Eq. \eqref{eq:4}, and we obtain
\begin{eqnarray}
 \frac{\partial n_{\boldsymbol{q},+}}{\partial t}&&= \frac{(I\sqrt{S}W_{+}(\boldsymbol{q}))^2}{2N}\sum_{\boldsymbol{k}}\bigg[
\delta(\epsilon_{\boldsymbol{k},\uparrow}+\omega_{\boldsymbol{q},+}-\epsilon_{\boldsymbol{k+q},\downarrow})\times \notag\\
&&\bigg(f_{\boldsymbol{k+q},\downarrow}(1-f_{\boldsymbol{k},\uparrow})-n_{\boldsymbol{q},+}(f_{\boldsymbol{k},\uparrow}-f_{\boldsymbol{k+q},\downarrow}) \bigg)\notag\\
    &&-\delta(\epsilon_{\boldsymbol{k},\downarrow}-\omega_{\boldsymbol{q},+}-\epsilon_{\boldsymbol{k+q},\uparrow})\times \notag\\
    &&\bigg(f_{\boldsymbol{k},\downarrow}(1-f_{\boldsymbol{k+q},\uparrow})-n_{\boldsymbol{q},+}(f_{\boldsymbol{k+q},\uparrow}-f_{\boldsymbol{k},\downarrow}) \bigg)\bigg], \label{eq:9}
\end{eqnarray}
for $\nu=+$. The case $\nu=-$ is obtained by exchanging $\uparrow$ and $\downarrow$. By solving the problems, see Appendix \ref{sec:A}, we can derive the expressions for the magnon spin currents
\begin{equation}
    {\bf J}^{\nu}_m=\frac{\hbar}{\Omega_m}\sum_{\boldsymbol{q}} {\bm\upsilon}_{\boldsymbol{q}}^{\nu}\partial_{\omega_{\boldsymbol{q},\nu}}n_{\boldsymbol{q},\nu}^{0} g^{\nu}_{\boldsymbol{q}},\label{eq:10}
\end{equation}
where $g^{\nu}_{\boldsymbol{q}}$ are given explicitly in the Appendix, ${\bm\upsilon}_{\boldsymbol{q}}^{\nu}$ is the speed of the magnons and $\Omega_m$ is the volume of the magnonic Brillouin zone. Finally, as a benchmark, we define the electron-mediated spin current for spin $\sigma$ as \cite{sinova2015spin}
\begin{equation}
   {\bf J}^{\sigma}_e = \frac{\tau_e e}{\Omega_e} \bigg(\frac{\hbar}{2}\bigg)\sum_{\bm k} {\bf v}^{\sigma}_{\bm k} \cdot ({\bf v}^{\sigma}_{\bm k}\cdot {\bf E}) \delta(\epsilon_{\bm k,\sigma}-\mu),
\end{equation}
where ${\bf v}^{\sigma}_{\bm k}$ is the electron velocity and $\Omega_e$ is the volume of the electronic Brillouin zone. As discussed above, in altermagnets, the electronic spin current ${\bf J}^{\uparrow}_e-{\bf J}^{\downarrow}_e$ exhibits MSHE \cite{gonzalez2021efficient}.

\section{MAGNON AND SPIN CURRENTS}\label{sec:iii}

We use the relation of the current to the electric field $J^{\nu,\sigma}_i=\sigma^{\nu,\sigma}_{ij}E_j$ to separate the conductivities of electron $\sigma$ and magnon $\nu$ along the directions $i,j=x,y$. By expressing the electric field direction with the angle $\theta$, the conductivities transverse and parallel to the electric field read
\begin{eqnarray}
\sigma_{\perp}^{\nu,\sigma}&=&(\sigma^{\nu,\sigma}_{xx}-\sigma^{\nu,\sigma}_{yy})\sin{\theta}\cos{\theta}+\sigma^{\nu,\sigma}_{xy}\sin^2{\theta}+\sigma^{\nu,\sigma}_{yx}\cos^2{\theta},\nonumber\\ \sigma_{||}^{\nu,\sigma}&=&\sigma^{\nu,\sigma}_{xx}\cos^2{\theta}+\sigma^{\nu,\sigma}_{yy}\sin^2{\theta}+(\sigma^{\nu,\sigma}_{yx}+\sigma^{\nu,\sigma}_{xy})\sin{\theta}\cos{\theta}.\nonumber
\end{eqnarray}
We define the total magnon current and magnon-mediated spin current as
\begin{eqnarray}
    \sigma^m_{\perp/||} &=& \sigma^+_{\perp/||}+\sigma^-_{\perp/||},\label{eq:11}\\
    \sigma^s_{\perp/||} &=& \sigma^+_{\perp/||}-\sigma^-_{\perp/||}. \label{eq:12}
\end{eqnarray}
Similarly, the total electron current and spin current read
\begin{eqnarray}
    \sigma^e_{\perp/||} = \sigma^{\downarrow}_{\perp/||}+\sigma^{\uparrow}_{\perp/||},\label{eq:11}\\
    \sigma^{s,e}_{\perp/||} = \sigma^{\downarrow}_{\perp/||}-\sigma^{\uparrow}_{\perp/||}. \label{eq:12}
\end{eqnarray}
Akin to the charge current in altermagnets \cite{gonzalez2021efficient}, the magnon current that flows transverse to the electric field, $\sigma^m_{\perp}$, vanishes as $\sigma^+_{\perp}=-\sigma^-_{\perp}$ for any variation of the parameters. The magnon current that flows parallel to the electric field, $\sigma^m_{||}$, depends on the chemical potential but not on the angle $\theta$, see also Appendix \ref{sec:A}. In other words, in such an altermagnet a longitudinal magnon flow always accompanies an electron flow, independently of the flow direction, as expected for conventional ferro- and antiferromagnets \cite{cheng2017interplay}. 

As already discussed, in the absence of an electric field, the two magnon modes are degenerate, even when electron-magnon interaction is turned on. In the presence of an electric field, the electron Fermi surfaces are imbalanced via the non-equilibrium dynamics of the electrons, as depicted in Figs. \ref{fig:fig3}(a) and (d) for two different angles of the electric field, $\theta=\pi/2$ and $\theta=\pi/4$. At $\theta=\pi/2$, the charge flow is not polarized, and therefore, both magnon chiralities are excited equally. Nonetheless, due to the anisotropy of the Fermi surface, magnons of opposite chirality flow at an opposite angle with respect to the charge flow, resulting in a magnonic MSHE. In contrast, at $\theta=\pi/4$, the anisotropy of the two Fermi surfaces results in a spin-polarized charge current, and therefore, one magnon chirality is more excited than the other, resulting in a net magnonic spin current flow along the charge current. In this case, no charge or magnonic MSHE is obtained \cite{gonzalez2021efficient}.

We first compute the electron-mediated spin conductivities transverse to and along the applied electric field in Fig. \ref{fig:fig3}(b) and (e), respectively. These conductivities, given as a function of the electric field angle $\theta$ and electron chemical potential $\mu$, exhibit the expected symmetries of the spin splitter effect \cite{gonzalez2021efficient}. In Fig. \ref{fig:fig3}(c) and (f), we compute the transverse and longitudinal magnon spin conductivities generated by the electron-magnon interaction. We find that the symmetry of the angle dependence of the magnon spin current conductivities is similar to the electron-mediated MSHE as explained above. Additionally, the spin conductivity vanishes when the chemical potential is outside the electron energy, as there are no electrons to scatter with magnons. A noticeable feature of the magnonic MSHE is its strong dependence as a function of the chemical potential, with sign reversal around, e.g., $\mu=\pm1$ eV. This feature is attributed to resonances at $\epsilon_{\boldsymbol{k},\uparrow}\approx\epsilon_{\boldsymbol{k+q},\downarrow}$ and $\epsilon_{\boldsymbol{k},\downarrow}\approx\epsilon_{\boldsymbol{k+q},\uparrow}$, see Eq. \eqref{eq:5}. Such resonances are similar to the one observed in electron-phonon interaction \cite{mahan2013many}. Close to these resonances, the magnon spin conductivity can reach about 10\% of the electron spin conductivity, which makes it detectable experimentally. 

\begin{figure}[t]
\includegraphics[width=8.6cm,height=8.6cm,keepaspectratio]{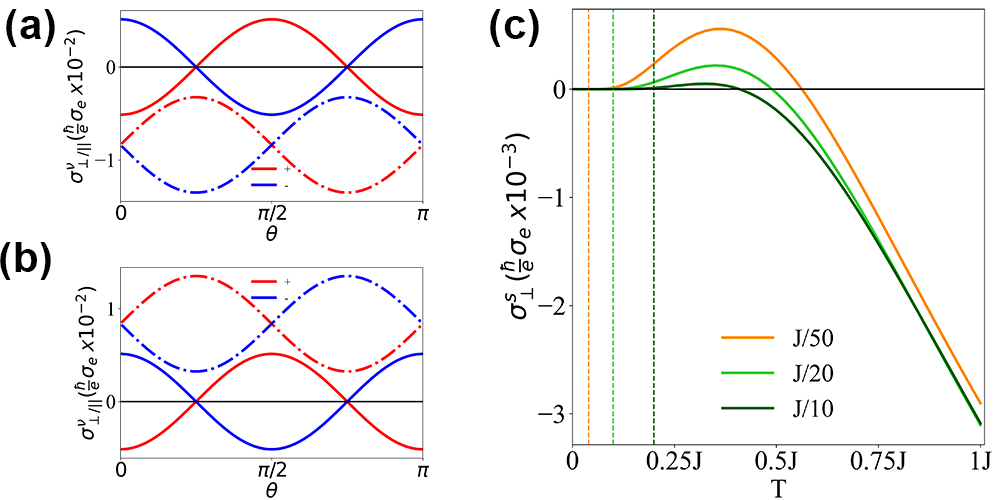}
\caption{(a,b) Angular dependence of the transverse ($\sigma^{\nu}_{\perp}$, solid lines) and longitudinal ($\sigma^{\nu}_{||}$, dashed lines) conductivities for the two magnon chiralities ($\nu=+1$ in red and $\nu=-1$ in blue). The chemical potential is set to (a) $\mu=+0.05$ and (b) $\mu=-0.05$. (c) Temperature dependence of the transverse magnon spin conductivity for different values of the easy axis anisotropy. The vertical dashed lines represent the activation temperatures.}
\label{fig:fig4}
\end{figure}

In Fig. \ref{fig:fig4}, we calculate the longitudinal and transverse magnon spin currents generated by the electron-magnon interaction of the two different magnon chiralities $\nu=\pm$. We can see that the transverse signal $\sigma^{\nu}_{\perp}$ is independent of the chemical potential near $\mu=0$. At the same time, the sign of the longitudinal conductivity $\sigma^{\nu}_{||}$ depends on the chemical potential and becomes maximum away from $\mu=0$, see Appendix \ref{sec:A}. In Fig. \ref{fig:fig4}(b), we plot the temperature dependence of the transverse magnon spin current for different values of the easy axis anisotropy ($\theta =0,\mu =0$).  At very low temperatures, i.e., below the magnon gap, equal to $2K$ in our model and indicated by the vertical dash lines, the magnon conductivity is zero and starts taking off only above it. Interestingly, we find that the magnon MSHE changes sign upon increasing temperature, a feature associated with the resonances pointed out above. Notice that such resonances and sign reversal depend on the specifics of the electron and magnon models, and realistic electronic band structure calculations are necessary to predict the temperature dependence of magnon-mediated MSHE. Upon further increasing the temperature, more magnons are thermally activated, leading to a linear increase in the magnon conductivity. Of course, at temperatures close to the Néel temperature, magnon softening leads to a collapse of the conductivity, not accounted for in our model.

\section{EXPERIMENTAL IMPLEMENTATION}\label{sec:iv}

We showed that when electrons flow in an altermagnet, they experience both electronic and magnonic MSHE. In addition, in the presence of spin-orbit coupling, ubiquitous in these materials, the SHE may develop that needs to be discarded from the data. To distinguish the transport of the two distinct particles, we draw attention to the fact that the temperature dependence of electrons is generally small due to their Fermi-Dirac statistics and large kinetic energies. Unlike electrons, magnons have a significant temperature dependence due to their lower energies and Bose-Einstein distribution. As depicted in Fig. \ref{fig:fig4}(b), the signal of the magnon spin currents should be zero for temperature below the magnon gap, where the magnons are not thermally excited and are activated as temperature increases. In contrast, the electron spin current should remain unchanged under such low-temperature changes. As such, the temperature dependence of the spin currents should give hints of its nature. The distinction between MSHE and SHE is more straightforward since the polarization of the spin currents induced by the former lies along the Néel vector, whereas that of the latter lies perpendicular to the plane.

To experimentally detect the magnon current and distinguish it from the electronic current, different setups can be used. A practical tool is the nonlocal detection scheme, widely used to 
study magnon propagation in insulators \cite{cornelissen2015long,lebrun2018tunable,xing2019magnon}. Whereas these studies only required two leads, the nonlocal detection of MSHE requires at least three contacts (one nonmagnetic injector, one nonmagnetic collector, and one magnetic collector), as achieved in Mn$_3$Sn by Kimata et al. \cite{Kimata2019}, which presents several challenges, among which the fact that the distance between the leads must be smaller than the (electronic and magnonic) spin relaxation length. Although no experimental estimate of these relaxation lengths exists in the literature, it is safe to speculate that such a distance should be smaller than a few tens of nanometers, making the fabrication daring. 

To circumvent this difficulty, an easier configuration consists in using magnetic multilayers in a current-in-plane configuration, as used to perform spin-orbit torque studies \cite{bose2022tilted,bai2022observation,karube2022observation,guo2024}, thermal spin injection \cite{bai2023efficient,liao2024separation}, and spin pumping \cite{wang2024inverse}. In a bilayer configuration, the ferromagnetic layer deposited on top of the altermagnet (possibly separated by a nonmagnetic spacer such as Cu) is used as a polarization detector that can discriminate between regular SHE and altermagnetic MSHE. Since we are interested in probing the temperature dependence of the incoming spin current, the spin-orbit torque phenomena would not be appropriate as it necessitates a large current flow accompanied by Joule heating. To avoid spurious heating effects, measurement the (possibly unidirectional) magnetoresistance response of the multilayer seems more appropriate \cite{Nakayama2013,Avci2015a}. As the magnon spin currents are highly sensitive to temperature contribution, experiments performed across a range of temperatures can help separate these contributions.

To our knowledge, an in-depth theoretical or experimental estimation of the magnon gap in altermagnetic candidate materials has yet to be made. Generally, the easy-axis anisotropy in the typical Heisenberg antiferromagnet can be in the range $K=J/100-J/10$. The Heisenberg exchange has been estimated as large as $100$ meV in $\mathrm{RuO_2}$ \cite{vsmejkal2022chiral}. This makes the temperature at which the spin current is sensitive to the thermally activated magnons at $T_s=11.6-116K$, which is well within the experimental range.

\section{CONCLUSION} \label{sec:v}

In this work, we have shown how to imprint the characteristics of the altermagnetic splitting of electrons to degenerate antiferromagnetic magnons and induce non-equilibrium magnonic spin currents. At equilibrium, the magnon sees the totality of the Fermi surface of the two electronic bands, which sums up to zero. Out of equilibrium, the picture is different; when the electric field excites electrons in one direction, the Fermi surfaces become imbalanced. The excited electrons scatter with magnons, generating transverse and longitudinal magnon spin currents on top of the electron spin current. The magnon spin currents inherit characteristics from both the electrons, i.e., the chemical potential dependence, and the magnons, i.e., the temperature dependence. This opens the way for generating and manipulating magnon spin currents via efficient methods that do not require elements with strong spin-orbit coupling. 


The present proposal raises several questions. First, a potential advantage of the magnonic MSHE, which remains to be proven experimentally, is the long spin relaxation length of magnons compared to that of electrons. If magnons enjoy a longer relaxation length than electrons, then the magnonic MSHE is expected to dominate over the electronic one. In addition, the present prediction is based on two simple models for electrons and magnons. In realistic materials, electronic structures are much more complicated due to the orbital character of the Bloch states, which might strongly impact the actual magnon-drag effect described in the work, especially via the electron-magnon resonances pointed out earlier. Finally, while our study assumed a degenerate magnon spectrum, the magnons may be chiral in these materials \cite{vsmejkal2022chiral,cui2023efficient,brekke2023two,Liu2024,Jost2025}, exhibiting the same symmetries as the electrons. Through electron-magnon interaction, each chiral magnon mode is dragged along the electric field, akin to the process depicted in Figs. 3(a) and (b), even in the absence of electron splitting. Therefore, a realistic treatment of the electron-magnon-induced MSHE should account for both electron and magnon splitting, which is likely to enhance the overall signal.

\acknowledgments
K. S. thanks Diego García Ovalle for fruitful discussions. K.S. and A.M. acknowledge support from the Excellence Initiative of Aix-Marseille Université–A*Midex, a French Investissements d’Avenir” program.

\appendix

\section{NON-EQUILIBRIUM MAGNONS} \label{sec:A}

In this Appendix, we present the explicit expressions of the nonequilibrium magnon distributions, following Ref. \cite{cheng2020spin}. By injecting the distribution functions of electrons [Eq. \eqref{eq:8}] and magnons [Eq. \eqref{eq:7}] in the definition of the magnon scattering [Eq. \eqref{eq:9}], we get the expression of $\partial n^{\nu}_{\boldsymbol{q}}/\partial t$ as a function of the equilibrium distributions $f^0_{\boldsymbol{k},\sigma}$ and $n^0_{\boldsymbol{q},\nu}$. To get rid of the energy derivatives of the distribution functions, we use the following identities 
\begin{eqnarray}
&&(1-f^0_{\boldsymbol{k},\sigma}+n^0_{\boldsymbol{q},\nu})\partial_{\epsilon_{\boldsymbol{k+q},\sigma'}}f^0_{\boldsymbol{k+q},\sigma'}=-\beta n^0_{\boldsymbol{q},\nu}f^0_{\boldsymbol{k+q},\sigma'}(1-f^0_{\boldsymbol{k},\sigma})\notag\\
   \\
&&(f^0_{\boldsymbol{k+q},\sigma'}+n^0_{\boldsymbol{q},\nu})\partial_{\epsilon_{\boldsymbol{k},\sigma}}f^0_{\boldsymbol{k},\sigma}=-\beta n^0_{\boldsymbol{q},\nu}f^0_{\boldsymbol{k},\sigma}(1-f^0_{\boldsymbol{k+q},\sigma'})\notag\\
   \\
&&\beta n^0_{\boldsymbol{q},\nu}=-\partial_{\omega_{\boldsymbol{q},\nu}}n^0_{\boldsymbol{q},\nu}/(n^0_{\boldsymbol{q},\nu}+1)
\end{eqnarray}
Then, this expression is injected in the Boltzmann transport equation, Eq. \eqref{eq:6}, which we solve to calculate the terms $g_{\bf q}^\nu$,

\begin{widetext}
\begin{eqnarray}
 &g^+_{\bm q}& =\frac{I^2S}{2N} \frac{(W^{+}_{\bm q})^2}{1+n^{+,0}_{\bm q}}\sum_{\bm k}\bigg[\delta(\epsilon_{\bm k,\uparrow}+\omega_{\bm q}^+-\epsilon_{\bm{k+q},\downarrow}) \bigg(f^0_{\bm k,\uparrow}(1-f^0_{\bm{k+q},\downarrow})\bigg)Q^1(\bm{k,q},\theta)\notag\\
    &+&\delta(\epsilon_{\bm k,\downarrow}-\omega_{\bm q}^+-\epsilon_{\bm{k+q},\uparrow})\bigg(f^0_{\bm k,\downarrow}(1-f^0_{\bm{k+q},\uparrow})\bigg)Q^2(\bm{k,q},\theta)\bigg]\bigg/\notag\\
    \bigg(&\frac{1}{\tau_+}&-\frac{I^2S}{2N} \frac{(W^{+}_{\bm q})^2}{1+n^{+,0}_{\bm q}}\sum_{\bm k}\bigg[
   \bigg(\delta(\epsilon_{\bm k,\uparrow}+\omega_{\bm q}^+-\epsilon_{\bm{k+q},\downarrow})(1-f^0_{\bm{k+q},\downarrow})f^0_{\bm k,\uparrow}-\delta(\epsilon_{\bm k,\downarrow}-\omega_{\bm q}^+-\epsilon_{\bm{k+q},\uparrow})(1-f^0_{\bm{k+q},\uparrow})f^0_{\bm k,\downarrow}\bigg)\bigg],\notag\\
   \quad\\
 &g^-_{\bm q}& =\frac{I^2S}{2N} \frac{(W^{-}_{\bm q})^2}{1+n^{-,0}_{\bm q}}\sum_k\bigg[\delta(\epsilon_{\bm k,\downarrow}+\omega_{\bm q}^--\epsilon_{\bm{k+q},\uparrow}) \bigg(f^0_{\bm k,\downarrow}(1-f^0_{\bm{k+q},\uparrow})\bigg)Q^2(\bm{k,q},\theta)\notag\\
    &+&\delta(\epsilon_{\bm k,\uparrow}-\omega_{\bm q}^--\epsilon_{\bm{k+q},\downarrow})\bigg(f^0_{\bm k,\uparrow}(1-f^0_{\bm{k+q},\downarrow})\bigg)Q^1(\bm{k,q},\theta)\bigg]\bigg/\notag\\
    \bigg(&\frac{1}{\tau_-}&-\frac{I^2S}{2N} \frac{(W^{-}_{\bm q})^2}{1+n_{\bm q}^{-,0}}\sum_k\bigg[
   \bigg(\delta(\epsilon_{\bm k,\downarrow}+\omega_{\bm q}^--\epsilon_{\bm{k+q},\uparrow})(1-f^0_{\bm{k+q},\uparrow})f^0_{\bm k,\downarrow}-\delta(\epsilon_{\bm k,\uparrow}-\omega_{\bm q}^--\epsilon_{\bm{k+q},\downarrow})(1-f^0_{\bm{k+q},\downarrow})f^0_{\bm k,\uparrow}\bigg)\bigg].\notag\\
   \quad
\end{eqnarray}
\end{widetext}

\begin{figure}[t]
\includegraphics[width=8.6cm,height=8.6cm,keepaspectratio]{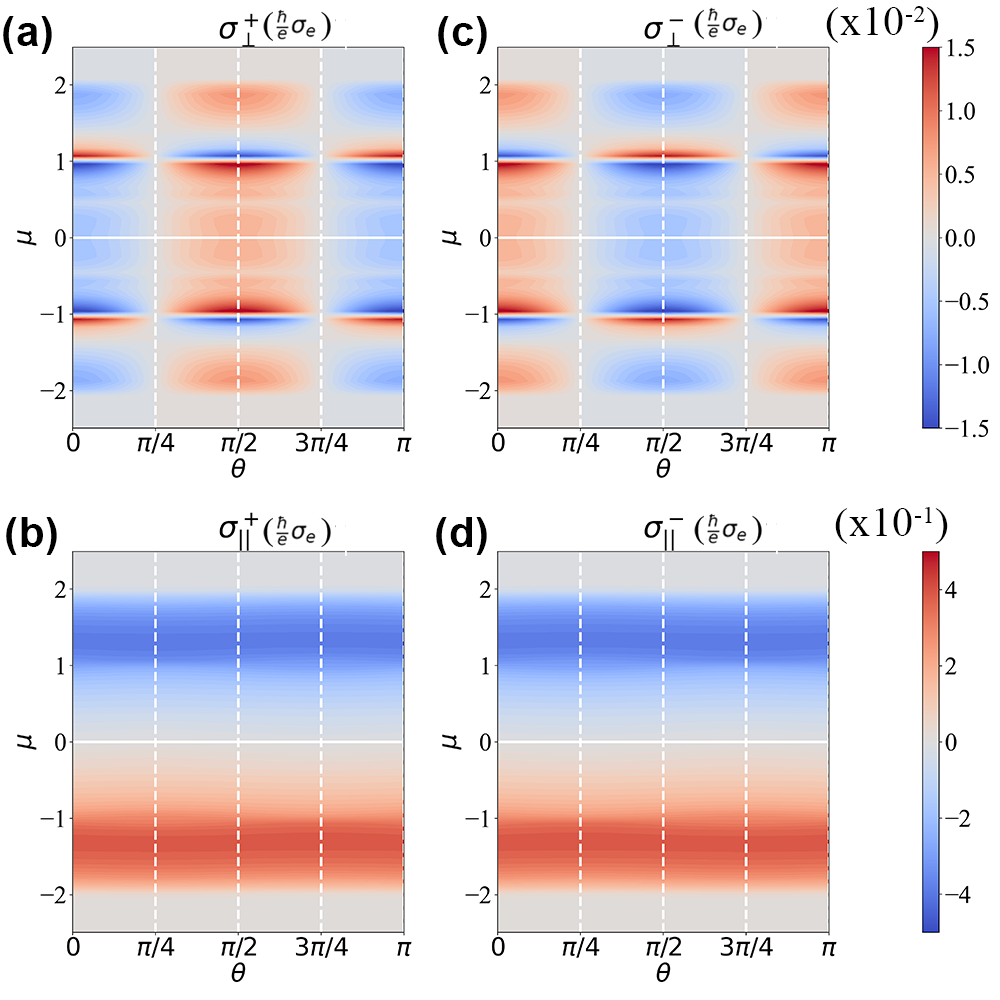}
\caption{The transverse and longitudinal magnon currents for the different magnon chiralities $\nu=\pm$. The parameters are the same as in the main text.}
\label{fig:figA}
\end{figure}

The expression of the electric field coupling to the electron can be written as
\begin{eqnarray}
    Q^1(\bm{k,q},\theta)=e\hbar\tau^{\uparrow}\bigg({\bf v}_{\bm k}^{\uparrow}\cdot {\bf E}\bigg)-e\hbar\tau^{\downarrow}\bigg({\bf v}_{\bm{k+q}}^{\downarrow}\cdot {\bf E}\bigg),\\
    Q^2(\bm{k,q},\theta)=e\hbar\tau^{\downarrow}\bigg({\bf v}_{\bm k}^{\downarrow}\cdot {\bf E}\bigg)-e\hbar\tau^{\uparrow}\bigg({\bf v}_{\bm{k+q}}^{\uparrow}\cdot {\bf E}\bigg),
\end{eqnarray}
and ${\bf v}^{\sigma}_{\bm k}=\partial \epsilon_{{\bm k},\sigma}/\partial {\bm k}$ is the velocity of the electrons. Throughout this work, we have considered the lifetimes of electrons $\tau^{\downarrow}=\tau^{\uparrow}=\tau_e$ and magnons $\tau^{+}=\tau^{-}=\tau_m$. The electron hopping integral can be related to its mass by setting $2t=1/m^*$, where $m^*$ is the effective electron mass in the metal. We use Drude's expression of the electron conductivity $\sigma_e$ to express the magnon spin conductivity in terms of $(\hbar/e)\sigma_e$.

In Fig. \ref{fig:figA}, we calculate the longitudinal and transverse magnon conductivities of the individual magnon chiralities, which are used to compute the net magnonic spin conductivities in Eqs. \eqref{eq:11} and \eqref{eq:12}. We note that the individual longitudinal conductivities are an order of magnitude larger than the transverse one. However, their net difference, the spin transport, is of the same order as the transverse one.

\bibliography{bibliography}

\end{document}